\documentstyle[11pt,newpasp,twoside]{article}
\def\edcomment#1{\iffalse\marginpar{\raggedright\sl#1\/}\else\relax\fi}
\reversemarginpar

\begin{document}
\title{Dark Matter and Dark Energy:  The Critical Questions}
 \author{Michael S. Turner}
\affil{Departments of Astronomy \& Astrophysics and of Physics,\\
Enrico Fermi Institute, and Center for Cosmological Physics\\
The University of Chicago, Chicago, IL~~60637-1433}
\affil{NASA/Fermilab Astrophysics Center, PO Box 500\\
Fermi National Accelerator Laboratory, Batavia, IL~~60510-0500}

\begin{abstract}

Stars account for only about 0.5\% of the content of
the Universe; the bulk of the Universe is optically dark.  The dark side of
the Universe is comprised of:  at least $0.1\%$ light
neutrinos; $3.5\% \pm 1\%$ baryons; $29\% \pm 4\%$ cold dark matter; and
$66\% \pm 6\%$ dark energy.  Now that we have 
characterized the dark side of
the Universe, the challenge is to understand it.
The critical questions are:  (1) What form do the
dark baryons take?  (2) What is (are) the constituent(s) of the cold dark matter?
(3) What is the nature of the mysterious dark energy that is causing the
Universe to speed up.

\end{abstract}

\section{Introduction}

The past five years have witnessed great progress in identifying
the basic features of our Universe.  It is spatially flat and
thus has the critical density ($\rho_{\rm crit} = 3H_0^2/8\pi G
\approx 10^{-29}\,{\rm g\,cm^{-3}}$).  The expansion is speeding
up, not slowing down (i.e., $q_0 < 0$).  The mass/energy density
is distributed as follows:

\begin{itemize}

\item Bright stars:  0.5\%
\item Baryons (total):  $4\%\pm 1\%$
\item Nonbaryonic dark matter:  $29\% \pm 4\%$
\item Neutrinos:  at least 0.1\% and possibly as large as 5\%
\item Dark Energy:  $66\% \pm 6\%$

\end{itemize}

The evidence for flatness comes from measurements
of the anisotropy of the cosmic microwave background (CMB) on
angular scales of about 1 degree.  The position of the first acoustic peak at
multipole number 200, as
determined by the BOOMERanG, MAXIMA, DASI and CBI experiments, implies
that $\Omega_0 = 1\pm 0.04$ (Sievers et al, 2002).
This means that the curvature radius of the Universe
is greater than about 5 times the Hubble radius since
$R_{\rm curv} = H_0^{-1}/|\Omega_0 -1|^{1/2}$.  In the next sections
I will discuss the evidence for the accounting of the various dark
components.

I did not mention the cosmic microwave background,
which today accounts for $\Omega_{\rm CMB} =  2.47h^{-2}\times 10^{-5}$
(about 0.005\%), or the relativistic
neutrino backgrounds which account for $\Omega_\nu = 0.56h^{-2}\times
10^{-5}$ per relativistic neutrino species.  Though unimportant to
the energy budget
today, at very early times relativistic particles were dominant.

The existence of (at least) three components that evolve differently
with redshift divides the evolution of the Universe into (at
least) three epochs:  1.  Early ($z> 10^4$ and $t< 10^4\,$yrs) 
radiation-dominated era (photons, neutrinos, and a spectrum of relativistic
particles that grows with temperature); 2. Matter-dominated era
($10^4 > z > 0.2$, $10\,{\rm Gyr} > t > 10^4\,$yrs) during which cosmic
structure grew; and 3. Dark-energy dominated era ($z< 0.2$, $t> 10\,$Gyr)
characterized by accelerated expansion and cessation of structure formation.

I note that the precision of the present accounting still
allows for an unidentified component that contributes perhaps as much
as 10\% of the critical density.

Now that we can enumerate the components of the Universe, the task is
to understand them.  Three critical questions arise:

\begin{enumerate}

\item What form do the dark baryons take?

\item What is (are) the constituent(s) of the nonbaryonic dark matter?

\item What is the nature of the dark energy?

\end{enumerate}

\section{Dark Baryons}

For many years the theory of big-bang nucleosynthesis (BBN) and a
lower limit to the primordial deuterium abundance were used to
argue for a low baryon density (see e.g., Schramm \& Turner, 1998),
$\Omega_B \la 0.1$.
The measurement of the primordial deuterium abundance in high-redshift
hydrogen clouds has turned the upper limit into the most precise determination of
the baryon density, $\Omega_B h^2 = 0.020\pm 0.001$, or $\rho_B =
3.8\pm 0.7 \times 10^{-31}\,{\rm g\,cm^{-3}}$ (Burles et al, 2001;
Tytler et al, 2000; O'Meara et al, 2001).
When combined with knowledge of the Hubble constant, $h=0.72\pm 0.07$
(Freedman et al, 2001), this implies
\begin{equation}
\Omega_B = 0.04 \pm 0.008
\end{equation}

Within the past year the credibility of the BBN baryon density got a
tremendous boost from the
DASI, BOOMERanG, MAXIMA, and CBI CMB experiments (Sievers et al, 2002).
Based upon the ratios of the
heights of the odd to even acoustic peaks, these experiments
imply:  $\Omega_Bh^2 = 0.022\pm 0.003$.
The agreement with BBN is striking, and the underlying physics
is very different:  gravity-driven acoustic oscillations at 400,000\,yrs vs.
nuclear physics at a few seconds.  Not only does this give one confidence
in a low baryon density, but it also provides a powerful consistency test
of the whole framework.

The absorption of light from $z\sim 3-4$ quasars by Lyman-alpha clouds
provides a lower limit to the fraction of critical density in
intergalactic gas which is consistent with BBN and the CMB:
$\Omega_Bh^2 > 0.018 [\Omega_Mh^2/0.17]^{1/4}$ (McDonald et al, 2001).
This determination of the baryon density depends upon a number of assumptions,
including the flux of ionizing radiation, the physics of the Lyman-alpha
forest being well described by CDM, and the bulk of the baryons
at $z\sim 3-4$ not being in stars or stellar remnants.

Today, baryons in stars account for only about 10\% of all baryons;
most of the baryons are still unaccounted for.  In rich
clusters, the census appears to be complete:  The ratio of baryons
in the hot, intracluster gas to that in stars is close to 10.  While clusters
provide an environment where it is easier to carry out a baryon census
because of their deep gravitational potential, clusters only account
for a few percent of the total matter density and a minority of the baryons.

Where then are the bulk of the baryons today?  The conventional
wisdom is that they reside in hot intergalactic gas or warm gas more closely
associated with galaxies (see e.g., Cen \& Ostriker, 1999).
While unlikely, it is possible that the bulk of
the baryons, or even a significant fraction,
exist as the remnants of an early generation of stars.

Finally, it is worth noting that at redshifts $z\sim 3, 1000, 10^{10}$
the baryon census is complete and consistent.  At BBN the baryons were in
neutrons and protons that were in the processing of being fused into the
light elements; at last scattering they were in ionized H and He gas; and at
$z\sim 3$, they were in intergalactic gas.
To complete the picture, we need a baryon census
today (see Fukugita et al, 1998 for a preliminary estimate).

\section{Nonbaryonic Dark Matter}

The total matter density can now be determined by measurements
that do {\em not} depend upon the connection between mass and light (see e.g.,
Turner, 2001).  They include CMB anisotropy (the heights of the acoustic peaks
determine $\Omega_Mh^2$ and $\Omega_Bh^2$), the power spectrum of
inhomogeneity (its shape determines $\Omega_Mh$ and $\Omega_M/\Omega_B$), the cluster
baryon fraction ($\Omega_M /\Omega_B$) and the primordial deuterium
abundance ($\Omega_Bh^2$).
From these a robust and ``unbiased'' value of the matter density follows (Turner, 2001):
\begin{equation}
\Omega_M = 0.33 \pm 0.035 \qquad \Omega_B = 0.004\pm 0.008 \ \ \Rightarrow\ \
\Omega_{NB} =0.29 \pm 0.04
\end{equation}
With a high degree of confidence, we can say that
almost 90\% of the matter is in a form other than
baryons and yet to be identified!  This conclusion receives support from another
important, though indirect, argument:  Absent nonbaryonic dark matter, there is
no model for structure formation that can begin with the level of matter inhomogeneity
indicated by the anisotropy of the CMB ($\delta \rho /\rho \sim 10^{-5}$) and
explain the structure seen in the Universe today.

For almost 20 years there has been a working hypothesis:  the bulk of the
dark matter exists in the form of stable (or longlived) relic particles
left over from the earliest moments of creation.  For almost as long,
the leading particle candidates have been:  one or more light neutrinos,
the axion and the neutralino.  Significant progress has been
and is being made toward testing all three candidates 
(see e.g., Turner, 2000a or Griest \& Kamionkowski, 2000).

\subsection{Neutrinos}

Results from the SuperKamiokande (SuperK, Fukuda et al, 1999; Fukuda et al, 2000)
and Sudbury Neutrino Observatory (SNO, Ahmad et al, 2002a,b) experiments
have provided strong evidence for neutrino oscillations, with two mass
differences identified:  $\Delta m_{12}^2 \simeq 10^{-4}\,{\rm eV}^2$ (SNO/solar
neutrinos) and $\Delta m_{23}^2 \simeq 3\times 10^{-3}\,{\rm eV}^2$
(SuperK/atmospheric).  Oscillation experiments only probe relative
masses, and not the absolute mass scale.  With the simple assumption of
a mass hierarchy, $m_3 \gg m_2 \gg m_1$, this
implies $m_3 \sim 0.05\,$eV and $m_2 \sim 0.01\,$eV,
corresponding to $\Omega_\nu \sim 0.1\%$ (note, $\Omega_\nu =
m_\nu/ 93.2h^2\,$eV).  Hierarchy or not, this provides
a lower limit to what neutrinos contribute to the mass budget -- in the
same ball park as stars -- and validates the concept of particle dark
matter.  Cosmology has long since made its judgment on neutrinos -- hot
dark matter cannot account for the structure we see today and even a
modest of amount of hot dark matter ($\Omega_\nu \ga 0.05$) leads to an
unacceptable deficiency of small-scale structure (Croft et al, 1998;
Elgaroy et al, 2002).

In setting the absolute scale of neutrino mass, cosmological observations
may play an important role.  CMB anisotropy and large-scale structure measurements
have sensitivity at the tenths of an eV level (see e.g., Hu et al, 1998).

\subsection{Cold dark matter}

Cosmology has also made its judgment on cold dark matter -- the
cold dark matter scenario has much, if not all, the truth about structure
formation.  The two leading CDM particles are the axion, a very light
particle ($m\sim 10^{-6}\,{\rm eV} -10^{-5}\,$eV), which is cold by virtue of being
born in a Bose condensate, and the neutralino, a heavy particle ($m \sim
50\,{\rm GeV} - 500\,$GeV) and the lightest of the supersymmetric partners of
the known particles, which is cold by virtue of its large mass.

Experiments to detect the axions
or neutralinos that may comprise our own halo have reached sufficient
sensitivity to begin seriously testing both possibilities.
In addition, the neutralino can be produced at an accelerator or detected
by its annihilation products (see e.g., Griest \& Kamionkowski, 2000).
(Neutralinos that accumulate in the
sun annihilate and produce high-energy neutrinos; neutralinos in
the halo can annihilate and produce photons, antiprotons or gamma rays.)

Like the neutrino, the interactions of the axion and neutralino with
ordinary matter are very weak (mean free path $\gg H_0^{-1}$).  So far
as cosmology goes, axions and neutralinos are particles that only
interact via gravity.  This fact has made simulation of
cosmic structure formation relatively simple (compared to the typical astrophysical
simulation).  Unfortunately, it also means that axionic cold dark matter
and neutralino cold dark matter cannot be distinguished by purely
cosmological observations.

There are some indications that CDM does not have all the truth
on small scales, namely, the halo cusp problem and the overabundance
of substructure (see e.g., Sellwood \& Kosowsky, 2000).
While plausible astrophysical explanations exist
for both problems (see, e.g., Benson et al, 2001;
Bullock et al, 2000; Klypin et al, 2001; Merritt \& Cruz, 2001; Milosavljevic
\& Merritt, 2001; Milosavljevic et al, 2001; Somerville, 2001; Weinberg \&
Katz, 2001), the solution could involve an
unexpected property of the dark-matter particle, e.g., large self-interaction
cross section (Spergel \& Steinhardt, 2000), large annihilation cross section
(Kaplinghat et al, 2000), or mass of around 1\,keV (i.e., warm dark matter).

At present, there is the tantalizing possibility that further
study of cosmic structure on small scales could reveal
a new property of the dark-matter particle; in any case, this
is the arena where CDM needs to be tested.

\section{Dark Energy}

Dark energy is my term for the causative agent of the current
epoch of accelerated expansion.  According to the second
Friedmann equation,
\begin{equation}
{\ddot R}/R = -4\pi G \left( \rho + 3 p \right) /3 ;
\end{equation}
thus, stress-energy with large negative pressure, can
produce accelerated expansion ($R$ is the
cosmic scale factor).

Dark energy has the following defining properties:
(1) it emits no light; (2) it has large, negative pressure,
$p_X \sim -\rho_X$; and (3) it is approximately
homogeneous (more precisely, does not
cluster significantly with matter on scales at least as large as clusters
of galaxies).  Because its pressure is comparable in magnitude
to its energy density, it is more ``energy-like'' than ``matter-like''
(matter being characterized by $p\ll \rho$).
Dark energy is qualitatively very different from dark matter.

\subsection{Two lines of evidence for an accelerating Universe}
Two independent lines of evidence point to an accelerating Universe.  The first
is the measurement of type Ia supernova light curves
by two groups, the Supernova Cosmology Project (Perlmutter et al,
1999) and the High-$z$ Supernova Team (Riess et al, 1998).
The teams used different
analysis techniques and different samples of high-$z$ supernovae
and came to the same conclusion:  the Universe is speeding up,
not slowing down.

The recent discovery of a supernovae at $z\simeq 1.76$ bolsters the case
significantly (Riess et al, 2001) and provides the first evidence for
an early epoch of decelerated expansion (Turner \& Riess, 2002).
SN 1997ff falls right on the accelerating Universe curve
on the magnitude -- redshift diagram,
and is a magnitude brighter than expected in a dusty open Universe.

The second, independent line of evidence for the accelerating Universe
comes from measurements of the composition of the Universe, which
point to a missing energy component with negative pressure (Turner, 2000b).
The argument goes like this.  CMB anisotropy measurements indicate that
the Universe is flat, $\Omega_0=1.0\pm 0.04$ (Sievers et al, 2002).
In a flat Universe, the matter density and energy density
must sum to the critical density.  However,
matter only contributes about 1/3rd
of the critical density, $\Omega_M = 0.33\pm 0.04$
(Turner, 2001), leaving two thirds of the critical density missing.

In order to have escaped detection this missing energy
must be smoothly distributed.
In order not to interfere with the formation of cosmic structure,
the energy density in this component must change
more slowly than matter, so that it was subdominant in the past.
For example, if the missing 2/3rds of critical density were smoothly
distributed matter ($p=0$), then linear density
perturbations would grow as $R^{1/2}$ rather than as $R$.  The shortfall
in growth since last scattering ($z\simeq 1100$) would be a factor of 30,
leading to far too little growth to produce the structure seen today.

The pressure associated with the missing energy component determines
how it evolves:
\begin{equation}
\rho_X \propto R^{-3(1+w)} \ \ \Rightarrow\ \ \rho_X /\rho_M \propto (1+z)^{3w}
\end{equation}
where $w$ is the ratio of the pressure of the missing energy component
to its energy density (here assumed to be constant).
Note, the more negative $w$, the faster the ratio of missing energy
to matter goes to zero in the past.  In order to grow the
structure observed today from the density perturbations indicated by CMB anisotropy
measurements, $w$ must be more negative than about $-{1\over 2}$
(Turner \& White, 1997), and since $q_0 = {1\over 2} + {3\over 2}w\Omega_X 
\sim {1\over 2} + w$, this implies $q_0<0$ and accelerated expansion.

\subsection{Gravity can be repulsive in Einstein's theory}

In Newton's theory mass is the source of the gravitational
field and gravity is always attractive.  In general relativity,
both energy and pressure source the gravitational field,
cf. Eq. 3.  Sufficiently large
negative pressure leads to repulsive gravity, and so accelerated
expansion can be accommodated.
Of course, that does not preclude that the ultimate explanation
for accelerated expansion lies in a fundamental modification
of Einstein's theory.

Repulsive gravity is a stunning feature of general relativity.
It leads to a prediction every bit as revolutionary as black holes  --
the accelerating Universe.  If the explanation for the accelerating
Universe fits within general relativity, it will be a major new triumph
for Einstein's theory.

\subsection{The biggest embarrassment in theoretical physics}

Einstein introduced the cosmological constant to balance the attractive
gravity of matter in order to create a static cosmological model.  He quickly
discarded the cosmological constant after the discovery of the expansion
of the Universe.

Quantum field theory makes the consideration of
the cosmological constant obligatory not optional.  The only
possible covariant form for the energy of the (quantum) vacuum,
$$ T_{\rm VAC}^{\mu\nu} = \rho_{\rm VAC}g^{\mu\nu},$$
is mathematically equivalent to the cosmological constant.
It takes the form for a perfect
fluid with energy density $\rho_{\rm VAC}$ and isotropic
pressure $p_{\rm VAC} = - \rho_{\rm VAC}$ (i.e., $w=-1$)
and is precisely spatially uniform.  Vacuum energy is the {\em almost}
perfect candidate for dark energy.

Here is the rub: the contributions of well-understood physics
(say up to the $100\,$GeV scale) to the quantum-vacuum energy
add up to $10^{55}$ times the present critical density.
This is the well known cosmological-constant
problem (see e.g., Weinberg, 1989; Carroll, 2001).

While string theory currently offers the best hope for
a theory of everything, it has shed precious little
light on the problem; in fact, it has raised some new issues.
The deSitter space associated with the accelerating Universe poses
serious problems for the formulation of string theory
(Witten, 2001).

The cosmological constant problem leads to a fork in the
dark-energy road:  one path is to wait for theorists to get the
``right answer'' (i.e., succeed in showing that indeed quantum vacuum
energy contributes $\Omega_\Lambda \simeq 2/3$); the other path is to
assume that even quantum nothingness weighs nothing and something else
with negative pressure must be causing the Universe to speed up.
Of course, theorists follow the advice of Yogi Berra:
where you see a fork in the road, take it.

\subsection{Parameterizing dark energy:  for now, it's $w$}

Theorists have been very busy suggesting all kinds of interesting
possibilities for the dark energy:  networks of topological defects,
rolling or spinning scalar fields (quintessence and spintessence),
influence of ``the bulk'', and the breakdown
of the Friedmann equations (Carroll, 2001; Turner, 2000a).

The uniformity of the CMB testifies to the near isotropy
and homogeneity of the Universe.  This implies that the
stress-energy tensor for the Universe must take the perfect
fluid form.  Since dark energy dominates
the energy budget, its stress-energy tensor must be to a
good approximation
\begin{equation}
{T_{X}}^\mu_\nu \approx {\rm diag}[\rho_X,-p_X,-p_X,-p_X]
\end{equation}
where $p_X$ is the isotropic pressure and the desired dark
energy density is
$$\rho_X = 2.8\times 10^{-47}\,{\rm GeV}^4$$
(for $h=0.72$ and $\Omega_X = 0.66$).  This corresponds to
a tiny energy scale, $\rho_X^{1/4} = 2.3\times 10^{-3}\,$eV.

The pressure can be characterized by its ratio to the
energy density (or equation-of-state):
$$w\equiv p_X/\rho_X$$
which need not be constant; e.g., it could be a function of $\rho_X$
or an explicit function of time or redshift.

For vacuum energy $w=-1$; for a network of topological defects
$w=-N/3$ where $N$ is the dimensionality of the defects (1 for
strings, 2 for walls, etc.).  For a minimally coupled, rolling scalar field,
$w$ is time dependent and can vary between $-1$ and $+1$.

I believe that for the foreseeable future
getting at the dark energy will mean trying
to measure its equation-of-state, $w(t)$.

\subsection{The Universe:  the lab for studying dark energy}

Dark energy by its very nature is diffuse and a low-energy
phenomenon.  It probably cannot be produced at accelerators;
it isn't found in galaxies or even in clusters of galaxies.
The Universe itself is the natural lab -- perhaps the only
lab -- in which to study it.

The primary effect of dark energy on the Universe
is on the expansion rate.  The first Friedmann equation
can be written as
\begin{equation}
H^2(z)/H_0^2  = \Omega_M(1+z)^3 +
    (1-\Omega_M) \exp \left[ 3\int_0^z\,[1+w(x)]d\ln (1+x) \right]
\end{equation}
where $\Omega_M$ is the fraction of critical density
contributed by matter today, a flat Universe is
assumed, and the dark-energy term follows
from energy conservation, $d(\rho_X R^3) = -p_X
dR^3$.  For constant $w$ the dark-energy
term becomes $(1-\Omega_M) (1+z)^{3(1+w)}$.  In a flat
universe $H(z)/H_0$ depends upon only two parameters:  $\Omega_M$ and $w(z)$.

While $H(z)$ is probably not directly measurable (however
see, Loeb, 1998), it does affect significantly two
observable quantities:  the (comoving) distance to an object
at redshift $z$,
\begin{equation}
r(z) = \int_0^z \,{dz\over H(z)},
\end{equation}
and the growth of (linear) density perturbations, governed by
\begin{equation}
\ddot\delta_k + 2H\dot\delta_k - 4\pi G\rho_M = 0,
\end{equation}
where $\delta_k$ is the Fourier component of comoving
wavenumber $k$ and overdot indicates $d/dt$.

The comoving distance $r(z)$ can be mapped out with the aid
of standard candles (e.g., type Ia supernovae)
by measuring luminosity distances $d_L(z) = (1+z)r(z)$.
It can also be probed by counting objects of a known intrinsic comoving
number density, through the comoving volume
element, $dV/dzd\Omega = r^2(z)/H(z)$.

Both galaxies and clusters of galaxies have been suggested
as objects to count (Newman \& Davis, 2000; Holder et al, 2001).
For each, their comoving number density evolves (in the case of
clusters very significantly).  However, it is believed that
much, if not all, of the evolution can be modelled through
numerical simulations and semi-analytical calculations.
In the case of clusters, evolution is
so significant that the number count test probe is affected
by dark energy through both $r(z)$ and the growth of density
perturbations, with the latter being the dominant effect.

The various cosmological approaches to ferreting
out the nature of the dark energy have been studied extensively.
Based largely upon my work with Dragan Huterer (Huterer \& Turner, 2001),
I summarize what we know about the efficacy of the cosmological
probes of dark energy:

\begin{itemize}

\item Present cosmological observations prefer $w=-1$,
with a 95\% confidence limit $w < -0.6$ (e.g., 
Perlmutter, Turner \& White, 1999).

\item Because dark energy was less important in the past,
$\rho_X/\rho_M \propto (1+z)^{3w}\rightarrow 0$ as $z
\rightarrow \infty$, and the Hubble
flow at low redshift is insensitive to the composition of
the Universe, the most sensitive
redshift interval for probing dark energy is $z=0.2 - 2$.

\item The CMB has limited power to probe $w$ (e.g.,
the projected precision for Planck is $\sigma_w = 0.25$)
and no power to probe its time variation.

\item A high-quality sample of 2000 SNe distributed from
$z=0.2$ to $z=1.7$ could measure $w$ to a precision $\sigma_w
=0.05$.  If $\Omega_M$ is known independently to better
than $\sigma_{\Omega_M} = 0.03$, $\sigma_w$ improves by
a factor of two and the rate of change of $w^\prime =
dw/dz$ can be measured to precision $\sigma_{w^\prime} = 0.16$.

\item Counts of galaxies and of clusters of galaxies may have
the same potential to probe $w$ as SNe Ia.  The
critical issue is systematics (including the evolution of
the intrinsic comoving number density, and the ability to identify
galaxies or clusters of a fixed mass).

\item Measuring weak gravitational lensing by large-scale
structure over a field of 1000 square degrees (or more)
could have comparable sensitivity to $w$ as type Ia supernovae.
However, weak gravitational lensing does not appear to be a good
method to probe the time variation of $w$ (Huterer, 2001).  The systematics
associated with weak gravitational lensing have not yet been studied
carefully and could limit its potential.

\item Some methods do not look promising in their ability
to probe $w$ because of irreducible systematics (e.g.,
Alcock -- Paczynski test and strong gravitational lensing
of quasars).  However, both could provide important independent
confirmation of accelerated expansion.

\end{itemize}

\end{document}